\documentclass[10pt,conference,a4paper,twocolumn]{IEEEtran}
\usepackage{cite}
\usepackage{amsmath,amssymb,amsfonts}
\usepackage{algorithmic}
\usepackage{graphicx}
\usepackage{textcomp}
\usepackage{xcolor}
\usepackage{url}
\usepackage{dblfloatfix}    

\usepackage{multicol,multirow,booktabs}

\usepackage{fancyhdr,lipsum}
\fancypagestyle{firstpage}{
\fancyhf{}
\fancyhead[C]{To appear at the 25th International Symposium on Design and Diagnostics of Electronic Circuits and Systems – DDECS 2022. Prague, Czech Republic.} 
\fancyfoot[C]{978-1-6654-9431-1/22/\$31.00~\copyright{}2022 IEEE \hfill}
}

\begin{document}
\title{\LARGE ArithsGen: Arithmetic Circuit Generator for Hardware Accelerators\vspace*{-0.5em}}

\author{
\IEEEauthorblockN{Jan Klhufek}
\IEEEauthorblockA{\textit{Faculty of Information Technology} \\
\textit{Brno University of Technology}\\
Brno, Czech Republic \\
xklhuf01@stud.fit.vutbr.cz\vspace*{-1cm}}
\and
\IEEEauthorblockN{Vojtech Mrazek}
\IEEEauthorblockA{\textit{Faculty of Information Technology} \\
\textit{Brno University of Technology}\\
Brno, Czech Republic \\
mrazek@fit.vutbr.cz
\vspace*{-0.8cm}
}

}

\maketitle
\thispagestyle{firstpage}

\begin{abstract}
Generators of arithmetic circuits can automatically deliver various implementations of arithmetic circuits that show different tradeoffs between the key circuit parameters (delay, area, power consumption). However, existing (freely-)available generators are limited if more complex circuits with a hierarchical structure and additional architecture optimization are requested. Furthermore, they support only a few output formats. In order to overcome the above-mentioned limitations, we developed a new generator of arithmetic circuits called ArithsGen. ArithsGen can generate specific architectures of signed and unsigned adders and multipliers using basic building elements such as wires and gates.  Compared to existing generators, the user can, for example, specify the type of adders used in multipliers. The tool supports various outputs formats (Verilog, BLIF, C/C++, or integer netlists). ArithsGen was evaluated in the synthesis and optimization of generic customizable accurate and approximate adders and multipliers. Furthermore, we used the circuits generated by ArithsGen as seeds for a tool developed to automatically create approximate implementations of arithmetic circuits. We show that different initial circuits (generated by ArithsGen) significantly impact the properties of these approximate implementations. The tool is available online at \textcolor{blue}{\url{https://github.com/ehw-fit/ariths-gen}}. 
\end{abstract}


\section{Introduction}

Modern hardware accelerators of neural networks, machine learning applications, or image processing typically employ an array of processing elements (PEs). A processing element is a simple unit performing a basic arithmetic operation such as addition or multiplication. Since there are many PEs in the accelerator, selecting the proper implementation of the arithmetic operation concerning area, delay, or power consumption is beneficial.

The standard automated synthesis tools select the implementation of the operation automatically. However, users can re-design its implementation, usually  at the RTL description (i.e., VHDL or Verilog). These languages are relatively cumbersome for this task, and it is not straightforward to propose generic and configurable architectures. Therefore, some automatic generators of arithmetic circuits were proposed. These tools receive the type of circuit (e.g., Wallace-tree multiplier), parameters (e.g., type of internal adder), and desired bit-width. The output is a synthesizable RTL description of the circuit.

Some of the existing generators \cite{Zimmermann98vhdllibrary,Watanabe-2008-ArithmeticMG,Automatic-Generation-System-Galois-Fields} suffer from two major problems: i) they are abandoned and they use unsupported technologies, or ii) they are not publicly available because the corresponding papers show only the main idea. Currently, there are only a few freely available tools capable of generating specific types of arithmetic circuits.  

\textit{Arithmetic Module Generator} \cite{Arithmetic-Module-Generator} represents an on-demand service for generating the circuits in Verilog/VHDL (HDL). Internally, it uses Arithmetic Circuit Graph (ACG) for a formal description, verification, and conversion of a circuit into a hierarchical HDL representation. This tool allows generating many types of circuits (adders, multipliers, multiply-and-accumulate circuits (MACs), multi-operand adders, etc.). Moreover, the user can specify implementation of internal adders, the representation of numbers (signed or unsigned), and the input word length. The tool shows the following issues: i) since the tool is web-based and on-demand, it is unfeasible to generate multiple circuits. ii) the tool is closed-source; therefore, the user cannot extend the supported circuit types. iii) only the final HDL description is available (no access to the internal ACG representation).

The tool called \textit{VHDLMultGenerator}~\cite{VHDLMultGenerator} uses a graphical user interface to allow the user to specify the desired architecture of multiplier. This application has closed source code. The output is limited to multipliers and VHDL representation. This tool represents a closed-source approach that cannot be extended.

Recently, \textit{GenMul} tool was released~\cite{MGD:2021b}. This tool allows to generate fully configurable signed and unsigned multipliers. The source codes are publicly available. It is possible to generate the multipliers in Verilog format only. The main advantage is a level of configurability of the multiplier. The user specifies the accumulator of partial products and the final stage adder. However, the scheme for generating the multipliers is hard-coded and it is very complicated to extend the tool to support different types of multipliers and other circuits.

\IEEEpubidadjcol

Based on the shortcomings mentioned above, we decided to implement a new generator. The generator is general, easily extensible, and supports basic optimizations. The resulting circuits are applicable not only in standard synthesis but also in some optimization or verification tools. These tools (like ABC~\cite{brayton:abc}) do not support HDL languages or only support their minimal subset. Therefore, besides standard Verilog/VHDL, we decided to use common Berkeley Logic Interchange Format (BLIF) and specialized integer netlist for Cartesian Genetic Programming~\cite{cgp}. 
Since some tools cannot work with a hierarchical structure, the proposed generator enables to flatten (ungroup) the hierarchy.
The generator can generate arbitrary combinational circuit, currently we support adders, multipliers, MACs, and approximate multipliers.

\section{ArithsGen generator specification}

Although HDL languages support generic constructions, converting the circuit to different formats supported in some tools (i.e., BLIF, integer netlist) is complicated. 
Despite the support for structural description in HDLs, arithmetic circuits contain regularities difficult to describe in standard notation. This is particularly evident in the case of a large number of parameters (e.g., various bit widths, selection of internal components or inner tree construction, etc.). Inheritance, argument branching, simpler condition handling, or exceptions supported in modern languages can help to create generic description.
We decided to implement a new, simple meta-language for structural description of combinational circuits. This language is implemented as a Python package because Python is class-oriented, supports inheritance, and has some useful constructions which make the circuit description clear. Using the basic building elements such as wires and logic gates, our proposed tool called ArithsGen can generate selected architectures of signed and unsigned adders or multipliers. These circuits are fully configurable. For example, the user can choose partial adders in multipliers.

\begin{figure}[b]
    \centering\vspace*{-1.3em}
    \includegraphics[width=0.9\columnwidth]{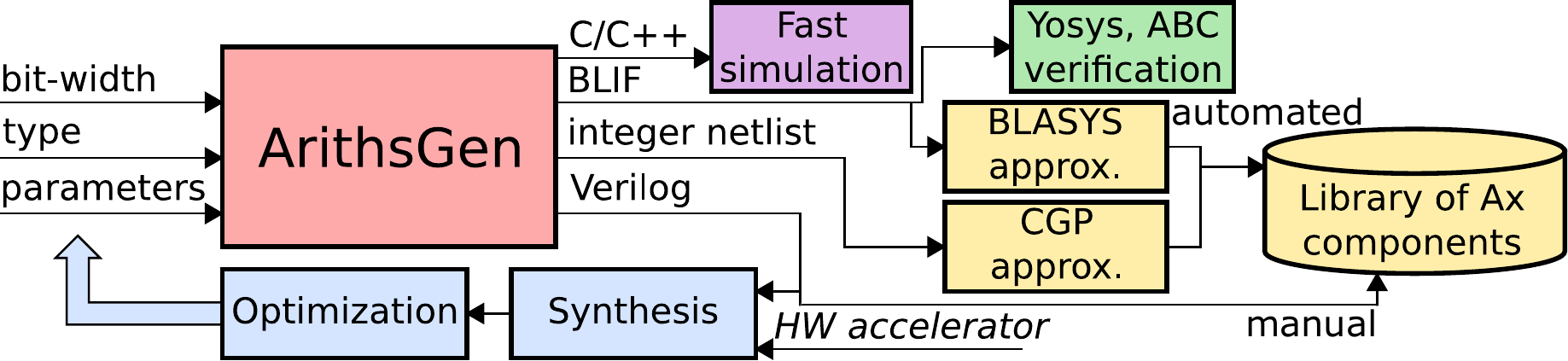}
    \vspace*{-1em}
    \caption{An example of use-cases of the proposed generator of arithmetic circuits.}\vspace*{-1em}
    \label{fig:overall}
\end{figure}

Figure \ref{fig:overall} shows the overall scheme of ArithsGen functionality. The ArithsGen module (i.e., Python object that uses the proposed framework) has configurable parameters (bit-width, type, additional parameters). The generator will particularly be helpful for the \textbf{design of HW accelerators} (blue boxes). The output Verilog code of the arithmetic circuit is the input for the synthesis and implementation process. Based on the synthesis results, some optimization algorithms can modify the configuration of ArithsGen to generate even better arithmetic circuits for the accelerator. Another use-case (green) of the proposed tool is \textbf{benchmarking of formal verification tools}. The BLIF code produced by ArithsGen can be used as a benchmark for different verification techniques~\cite{mahzoon:verification}. For \textbf{fast function verification} of the generated circuit, ArithsGen enables to generate the output file in C/C++. After compiling this output, the simulation is several orders of magnitude faster than the RTL level. The last envisioned use-case is the approximation of digital arithmetic circuits~\cite{mittal}. Some automated approximation methods use BLIF files \cite{hashemi2018blasys} or integer netlist known in Cartesian Genetic Programming (CGP)~\cite{ceska:iccad17}. The flat accurate circuit generated by ArithsGen can serve as a seed for these methods. 
Alternatively ArithsGen even enables to generate some manually designed approximate circuits natively. The users thus can create \textbf{approximate circuits} showing various tradeoffs between the error and other parameters (yellow boxes). The proposed generator ArithsGen is available as an open-source application including detailed tutorial at \textcolor{blue}{\url{https://github.com/ehw-fit/ariths-gen}}.

\section{Proposed methodology}
The proposed library ArithsGen employs Python's object orientation to model the modularity of arithmetic circuits and to describe their internal structures. We can divide the library into two parts: 1) core components such as gates, wires, and buses, and 2) arithmetic circuits. The structure of each component, starting with wires, buses, and logic gates, is described using its own class. Instances of classes representing the lower building components are used to describe more complex circuits. The appropriate wires of the circuit's subcomponents are then interconnected using their input-output interfaces. Having all this combined allows for the definition of numerous circuits that use previously defined circuits as their building components. Therefore, we can generate not only the basic arithmetic components but also complex circuits (i.e., MAC, DCT, etc.). On top of that, ArithsGen also focuses on exporting the circuit from its meta-language representation into various other output representations for further work such as C/C++, Verilog, BLIF, and the integer netlist for CGP.

\begin{figure}[!b]
    \centering
    \vspace{-6mm}
    \includegraphics[width=\columnwidth]{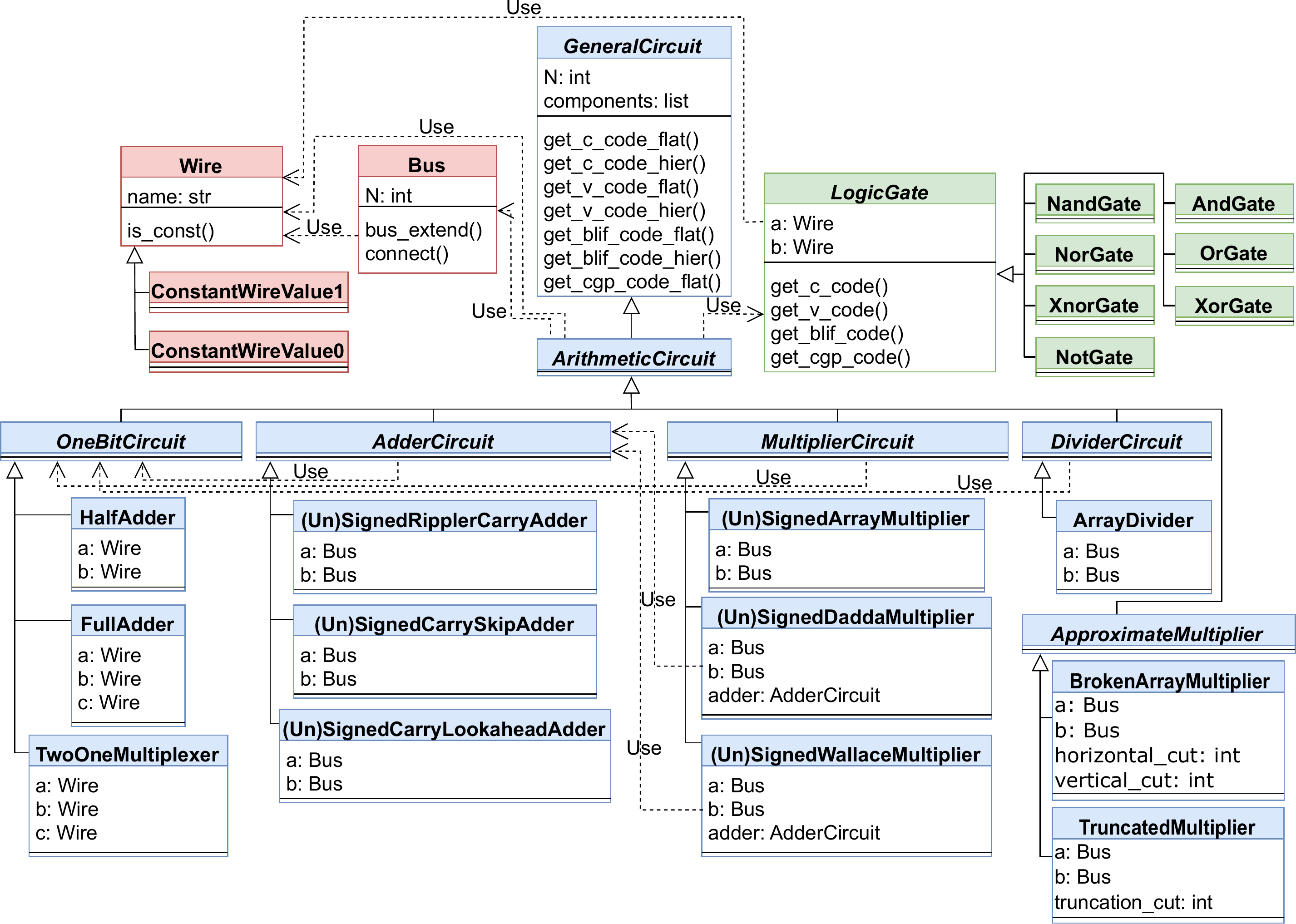}
    \vspace*{-8mm}
    \caption{Class diagram of the proposed \textit{ArithsGen} library. The closed arrow describes that the class inherits (implements) the referenced class. The open arrow shows that the class uses the referenced ones.}
    \vspace*{-1.4em}
    \label{fig:arithsgen_classes}
\end{figure}

\subsection{Wires and buses}
Single-bit wires and multi-bit buses are used to interconnect circuits through their I/O interfaces. Buses are defined by their bit width and the wires they contain. In addition to the general wire, there are also definitions of classes for constant wires, i.e., the wires that are connected either to a voltage source or to the ground. The red components on the left in Figure \ref{fig:arithsgen_classes} depict the logic of the described classes.

\subsection{Logic gates}
Logic gates are the most basic logically separate functional units from which more complex circuits are created. Depending on the type of input wires (general/constant wire) and their combination, the structure of the gate can be simplified or omitted for the generation to achieve internal optimization of the circuit design. Advanced design optimization will then be carried out by the synthesis tool applied on the output of ArithsGen. In Figure \ref{fig:arithsgen_classes}, the green components on the right show the overview of ArithsGen logic gates classes and their linkage to previously defined wire components.

\subsection{Arithmetic circuits}
The basic definition of arithmetic circuits is found in the abstract classes, located on the highest level of the class hierarchy. Thanks to the use of inheritance, this allows for a uniform way of defining different types of arithmetic circuits independent of their bit width (one-bit, multi-bit), type (adder, multiplier), architecture (e.g., RCA, Dadda), number representation system (signed, unsigned) and much more. To get a better idea about the proposed concept, the blue components in the middle of Figure \ref{fig:arithsgen_classes} show the distinct subclasses of arithmetic circuits.

\subsubsection{One bit circuits}
One bit circuits represent components that use wires as inputs and whose internal structures are built from the appropriate gates objects and their interconnection. This type includes, for example, a half or a full adder.

\subsubsection{Complex circuits}
The complex circuits are built from basic one-bit components and recursively from other complex circuits. We defined a set of six variable signed and unsigned adders (RCA, CSkA, and CLA), while each is represented as a class that inherits the \texttt{ArithmeticCircuit} definition. The definitions of individual adder architectures consist of sequential addition and interconnection of single-bit subcomponents. 
The users can define their own arithmetic circuit just by the structural description. For more details please follow the tutorial in the repository.

Multiplier architectures work with more complex logic than adders due to their need to generate an array of partial products and their subsequent accumulation. Therefore, the multiplier superclass contains methods that are used to generate and reduce partial products. Besides the sequential addition and interconnection of single-bit subcomponents used to define them, some multipliers such as Dadda and Wallace also allow the parametric selection of an unsigned multi-bit adder used to sum the reduced bits of partial products. The user can specify an arbitrary adder (even some new ones) by specifying the \texttt{unsigned\_adder\_class\_name} parameter in the constructor.

The currently implemented approximate multiplier architectures are broken-array multiplier (BAM) and truncated multiplier (TM). Their design corresponds to an array multiplier with some ommited partial product cells to introduce an error.

\begin{figure}[th]
    \centering
    \includegraphics[width=0.9\columnwidth]{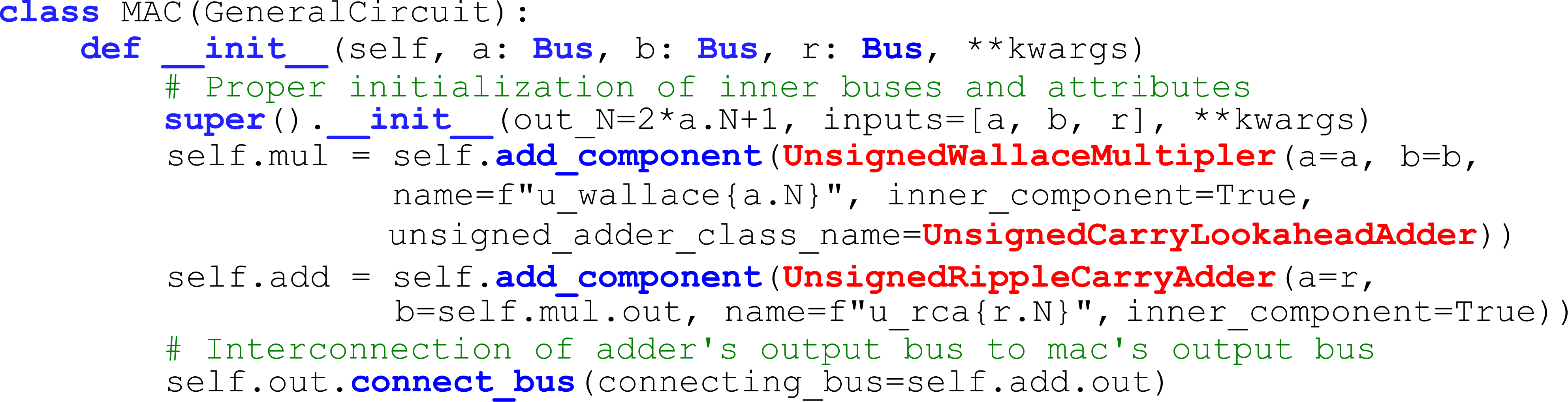}
    \vspace{-1em}
    \caption{An example code of an unsigned multiply-and-accumulate circuit that consists of $n$-bit array multiplier and $2n$-bit RCA. The red sections signify the parametric choice of adder and multiplier circuits that can be selected by means of an optimization algorithm.
    An advanced level of configurability could be introduced through the class instantiation. It is also possible configure ArithsGen to use specialized PDK components (e.g., full-adders, MUXes).} \vspace*{-1.5em}
    \label{fig:mac_code}
\end{figure}

To demonstrate different types of arithmetic circuits, we also implemented an Array divider based on a series of iterative subtractions. In Figure~\ref{fig:mac_code} we show the capability of the framework to construct complex components from previously defined building blocks. The code represents a widely-used multiply-and-accumulate circuit that performs $(a\cdot b) + r$ operation. Similarly, more complex arithmetic circuits such as logarithmic or floating-point  multipliers could be added. The interconnection, flattening, and export from the inner meta-language representation to the desired output formats are natively supported by the ArithsGen framework.

\subsection{Output formats}
The design generated using ArithsGen can be exported from its inner representation into one of the following formats: Verilog, BLIF, C/C++ and integer netlist for CGP. The export is performed by a way of recursively extracting information about all the various subcomponents hierarchically composing the exporting circuit. Since some tools do not support hierarchical structure, ArithsGen implements flattening --- the subcomponents are recursively extracted to the level of the basic operations (gates). The framework must guarantee that the names of wires are unique in each instance of the component. For the flat format, we also support the integer netlist used by tools for circuit optimization and approximation based on CGP. The code is generated in the following steps: i) creating the input buses, ii) instantiating the selected arithmetic circuit, and iii) calling the proper function (e.g., \texttt{get\_verilog\_code\_\{flat|hier\}}) to get the desired representation. Note that all of the output representations except for integer netlist support both the flattened and hierarchical export variants. The flattened netlist is used to represent interconnections between the simplest components -- gates along with all their connectivity information preserved.

\begin{figure*}[t!]
    \centering
    \vspace*{-1em}
    
    \includegraphics[width=\textwidth]{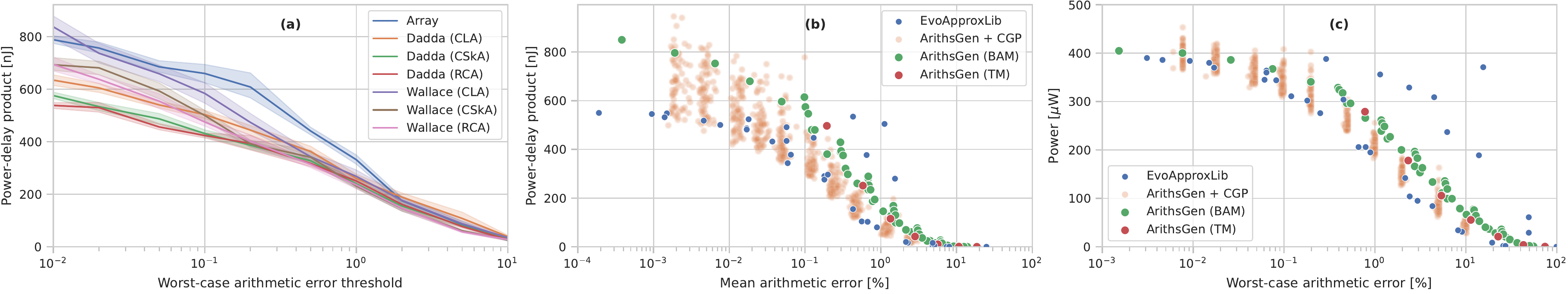}
    \vspace*{-2em}
    \caption{Results of approximation of arithmetic circuit \textbf{(a)} The final PDP of approximate 8-bit multipliers constructed from different seeds for a given maximal worst-case error. The filled area shows a 0.95 confidence interval; \textbf{(b,c)}  Parameters of resulting circuit approximated using CGP with different seeds compared to manually designed Broken-array (BAM) and Truncated (TM) multipliers \textit{(both implemented in ArithsGen)}, and EvoApproxLib~\cite{mrazek:date16lib}. }
    \vspace*{-1.5em}
    
    \label{fig:ax_res}
\end{figure*}

\section{Experiments}

\subsection{Validation and verification}
In the first experiment, we verified the validity of the generated arithmetic circuits. We used simulation tools to ensure correct functionality of smaller circuits and ABC verification for bigger circuits. 
We also observed that 32-bit circuits are generated in less than 0.5 seconds for all output formats (e.g., up to 12,094 lines of code for flat 32-bit multiplier).

\subsection{Scenario I: Hardware accelerator design}
The first experimental scenario focuses on the design of hardware accelerators. We evaluated the constructed arithmetic circuits in ASIC design (Synopsys Design Compiler and FreePDK45 library, optimized for delay with high effort). We investigated that flat adders have 25-31\% better power consumption than hierarchical adders. This was caused by the fact that the synthesis tool can better optimize the flat structure. Please note that the synthesis tool supports flattening natively. However, there were no significant power differences for very large multipliers after flattening.

To demonstrate the impact of ArithsGen setting on properties of the selected multiplier, we evaluated 16-bit multipliers in detail for some of the synthesis parameters (Table~\ref{tab:mul16}). Note that implementation parameters influencing the final layout were not considered. We observed that Dadda multipliers achieved 10\% area savings in contrast with the array multiplier. However, the latency is the same or even worse (by 4\%) for the signed variant. Dadda multipliers improve the power consumption by 14-23\%. The Wallace-tree multipliers have the worst array and delay, but they have shown excellent power consumption. For the partial product and final stage adder, it is better to use RCA or CSkA than CLA adders.

\begin{table}[t]
    \centering
    \caption{Parameters of synthesized 16-bit multipliers}
    \label{tab:mul16}
    \vspace{-1em}
    \resizebox{\columnwidth}{!}{
    
    \begin{tabular}{c|cc|cc|cc}\toprule
        \multirow{2}{*}{\bf Multiplier Type} & \multicolumn{2}{c|}{\textbf{Area} [$\mu m^2$]} & \multicolumn{2}{c|}{\textbf{Delay} [ns]} & \multicolumn{2}{c}{\textbf{Power} [mW]} \\
    
         & Unsign. & Sign.~ & Unsign. & Sign.~ & Unsign. & Sign.~ \\\midrule
         Array 	&	5258	&	5311	& \bf	5.63	&	5.43	&	3.87	&	3.82\\\midrule
Dadda  (CLA)	&	4951	&	4856	&	5.65	&	5.45	&	3.10	&	3.28\\
Dadda  (CSkA)	&  4902	&	4816	&	5.86	& \bf	5.39	&	2.98	&	3.17\\
Dadda  (RCA)	&	\bf 4902	&	\bf 4815	&	5.87	&	5.40	&	2.98	&	3.17\\\midrule
Wallace tree  (CLA)	&	5643	&	5692	&	5.78	&	6.04	&	3.05	&	3.20\\
Wallace tree  (CSkA)	&	5531	&	5539	&	5.81	&	5.87	&	\bf 2.93	&	\bf 3.07\\
Wallace tree  (RCA)	&	5532	&	5537	&	5.81	&	5.87	&	2.93	&	3.07\\\bottomrule
    \end{tabular}
    }
    \vspace{-2em}
\end{table}

\subsection{Scenario II: Approximation of arithmetic circuits}
In the second experiment we approximated accurate 8-bit circuits to achieve a good worst-case arithmetic error (WCE)\footnote{Defined as $\max_{\forall x \in B^{n}}|f(x) - f'(x)|$, when $f$ is accurate and $f'$ is approximate function, and $B^{n}$ is a set of all possible input vectors.} and area trade-off. We decided to use Cartesian genetic programming approximation.  This algorithm starts with an accurate circuit (the so-called parent) and iteratively modifies it. The key idea is that the algorithm accepts the random modification as a new parent in the next step if and only if the area is better or equal to the current parent, and the WCE is below the given threshold. For more details about this algorithm, please follow~\cite{mrazek:iccad16,ceska:iccad17}. Since the CGP algorithm is non-deterministic, we run ten independent runs for each configuration and limited each run to 2 hours. Please note this experiment would not have been possible without the proposed generator of seeds in a specific integer flat format.

Figure~\ref{fig:ax_res}a shows the final power-delay product (PDP) for various seeds and WCE thresholds. We can see that Wallace-tree (CLA) is almost always the worst seed, but for larger WCEs the results are competitive to the others. On the other hand, Dadda (RCA) is the best seed to reach small WCEs, but less competitive for increased WCE. Note that the approximation method is always the same, only its seed is modified in our experiments.

We also compared the resulting circuits with the very last version of the state-of-the-art library of approximate components EvoApproxLib~\cite{mrazek:date16lib}. This library was constructed by CGP with one accurate circuit as a seed. We also implemented a generic description Broken-Array and Truncated multiplier, widely used configurable manually designed multipliers \cite{jiang}, in ArithsGen. Different configurations were also compared to the resulting circuits. Figures~\ref{fig:ax_res}bc show that the different seeds allow for improving PDP vs. Mean-arithmetic error trade-off significantly. Although the library was constructed for power as the main design objective, the proposed results slightly improve power vs. WCE trade-off too.

\section{Conclusion}
In this work, we introduce a new generator of arithmetic circuits. In contrast to existing solutions, the proposed tool supports multiple output formats (even the flattened one), is easy-to-extend, and is open-source. ArithsGen also allows the construction of complex circuits such as MAC. In addition to the standard application in hardware accelerators, this tool enables the users to benchmark verification algorithms or improve libraries of approximate circuits.

{
\noindent\textit{Acknowledgement} This work was supported by the Czech science foundation project GA22-02067S.
}
\vspace{-1mm}

\bibliographystyle{./bibliography/IEEEtran}
\bibliography{arithsgen,./bibliography/IEEEabrv,./bibliography/IEEEexample}

\end{document}